\documentclass[useAMS,usenatbib]{mn2e}
\usepackage{graphicx,graphics}

\newcommand{\arcs}{$^{\prime\prime}$}

\title[A slowly rotating bar in the dwarf irregular NGC 3741]{ A slow bar in the dwarf irregular galaxy NGC 3741 }
\author[A. Banerjee, N. N. Patra, J. N. Chengalur and A. Begum]{Arunima Banerjee\thanks{E-mail:
arunima@ncra.tifr.res.in (A BANERJEE); narendra@ncra.tifr.res.in (NNP); chengalur@ncra.tifr.res.in (JNC);
ayesha@iiserb.ac.in(A BEGUM)}, 
Narendra Nath Patra\footnotemark[1], Jayaram N. Chengalur\footnotemark[1] 
and \newauthor Ayesha Begum\footnotemark[2]\\
\footnotemark[1]National Centre for Radio Astrophysics, TIFR, Pune - 411007, India\\
\footnotemark[2]Indian Institute of Science Education and Research, Bhopal - 462023, India \\ }
\begin{document}



\maketitle

\label{firstpage}

\begin{abstract}

Using the Tremaine-Weinberg method, we measure the speed of the HI  bar seen in 
the  disk of NGC 3741. NGC 3741 is an extremely gas rich galaxy with an 
\mbox{H\,{\sc i}} disk which extends to about 8.3 times its Holmberg radius. 
It is also highly dark matter-dominated. Our calculated value of the pattern 
speed $\Omega_p$ is 17.1 $\pm$ 3.4~km $\textrm{s}^{-1}\textrm{kpc}^{-1}$. 
We also find the ratio of the co-rotation radius to the bar semi-major axis 
to be (1.6 $\pm$ 0.3), indicating a slow bar.  This is consistent with 
bar models in which dynamical friction results in a slow bar in dark matter 
dominated galaxies.

\end{abstract}


\section{Introduction}

Understanding the dynamics of galactic bars and its implications on galaxy evolution continues to be an important aspect of studies
of galaxy evolution. Bars are found to be present in about half of all disk galaxies (Binney \& Tremaine 2008). 
Galactic bars regulate angular momentum transfer, help in gas inflow from the outer galaxy, and thus, in fueling nuclear star formation (Martinet \& Friedli 1997; Aguerri 1999; Ellison et al. 2011). 
Bars may also drive spiral structure and result in disk heating (Saha et al. 2010), thereby playing a pivotal role in the secular evolution of the galactic disk (See Sellwood \& Wilkinson 1993 for a review).  
 Theoretical and kinematical studies indicate that the bar pattern rotates as a whole with a constant speed known as the bar pattern speed. Measurement of the pattern speed serves as a useful diagnostic probe of the structure and dynamics of the bar.
In the last few decades, several indirect techniques have been developed to estimate the pattern speed  (See Corsini (2008) for a review). A particularly interesting direct method was proposed by Tremaine \& Weinberg (1984). This is based on the observed
photometry and kinematics of a suitable tracer population, suitable in the sense that the tracer population obeys the continuity equation.
The Tremaine-Weinberg (TW) method has been used to measure the pattern speeds of several barred galaxies (Merrifield \& Kuijken 1995; Gerssen et al. 1999;
Bureau et al. 1999; Debattista et al. 2002; Aguerri et al. 2003; Gressen et al. 2003; Debattista \& Williams 2004; Hernandez et al. 2005;
Emsellem et al. 2006; Corsini et al. 2007; Chemin \& Hernandez 2009; Speights \& Westpahl 2012).

The pattern speed is known to correlate with other parameters of the host galaxy although the reasons for those are not very
well understood. For example, the pattern speed is observed to be correlated with the disk morphology, with late-type galaxies hosting slower bars in general. (Combes \& Elmegreen 1993; Rautiainen et al. 2008; Buta \& Zhang 2011). Further, dynamical friction due to a dense dark matter halo is expected to slow down the bar (Debattista \& Sellwood 1998, 2000; Athanassoula 2003; Valenzuela \& Klypin 2003), but unambiguous evidence for this is  still lacking.
Finally, although both tidal interactions and the galactic environment 
are expected to play a role in bar formation and its  pattern speed,
the exact details are not well understood (but see Miwa \& Noguchi 1998; M'ndez-Hern'ndez et al. 2011). 

Observational studies indicate that bars in general are ``fast" i.e., 
the ratio $\cal R$ of the corotation radius $R_{cr}$ to the bar semi-major axis $r_b$ is greater than 1.4 (See \S 4.2). But these studies 
have been biased towards the large, high surface brightness (HSB) galaxies in which the role of the dark matter may not be so important.
The few exceptional cases where ``slow" bars are detected are the dark matter dominated
blue compact dwarf galaxy NGC 2915 (Bureau et al. 1999) and the low surface brightness (LSB) galaxy UGC 628 (Chemin \& Hernandez 2009).
The presence of slow bars in these galaxies may be
attributed to the deceleration due to the dynamical friction of the dense dark matter halo as predicted earlier by hydrodynamical N-body 
simulation studies of galaxies. However the number of available measurements is too small to arrive at any firm conclusion.
Measuring the bar pattern speed, in more dark matter rich galaxies, is crucial to assess the role played by the
dark matter in regulating the bar dynamics as well as the evolutionary history of the disk as a whole.

In this paper, we apply the TW method to measure the pattern speed of the dwarf irregular galaxy NGC 3741. NGC 3741 is  
very gas rich with $M_{\mbox{H\,{\sc i}}}/L_B = 6.26 $ (Begum et al. 2008a) and marked by the presence of a giant gas disk and purely gaseous substructures like a bar and a spiral arm. 
This is interesting in itself since there are very few
examples of purely gaseous bars and spiral arms. To our knowledge, 
NGC 2915 (Bureau et al. 1999) is the only other  galaxy that has been found to host such purely gaseous substructures. Spleights \& Westphal (2012)
study several gas-rich galaxies but none of them appear to host purely gaseous ``classical" bars and spiral arms which have a well-defined pattern speed.
Also, NGC 3741 is dark matter dominated at all radii including the galactic centre where the bar is located (Begum et al. 2005, Gentile et al. 2007) and therefore 
provides the ideal site to study the effect of dark matter dynamical friction on the bar pattern speed.  

The rest of this paper is organized as follows. In \S 2, we describe the TW method, in \S 3  we give a brief overview of NGC 3741 and discuss the applicability of the TW method
to its pattern speed determination with \mbox{H\,{\sc i}} as tracer, in \S 4  we present our calculations and results followed by discussion 
and conclusions in \S 5 and \S 6 respectively.

\section[]{The Tremaine-Weinberg Method}

Tremaine \& Weinberg (1984) proposed a direct method of measuring the pattern speed of a galaxy employing the observed surface brightness and kinematics of a tracer population,
which is assumed to obey the equation of continuity. For such a tracer population, the pattern speed $\Omega_p$
is given by
$$ \Omega_p \sin(i)= \frac {\int\limits_{-\infty}^\infty \Sigma(x,y)[{\bar{v}}_{los}(x,y) - v_{sys}]dx } {\int\limits_{-\infty}^\infty \Sigma(x,y)x dx} \eqno (1a)$$
where  
``$i$" is the disk inclination,
$\Sigma(x,y)$ is the surface density of the tracer,  
``${{\bar{v}}_{los}}$" the mean line-of-sight velocity,
``$v_{sys}$" the systemic velocity of the galaxy,
and ``$x,y$" the cartesian co-ordinate system in the plane of the sky 
with the origin coinciding with the centre of the galaxy and the $x$-axis parallel to the 
line of nodes i.e. the line of intersection of the sky plane and galaxy plane.
It is assumed that the galactic disk 
is not warped.
The main advantage of the TW method is that it is a kinematic method and independent of any underlying 
dynamical model for the galaxy. 

Equation (1a) is generally rewritten as follows. 
Let ${\int\limits_{-\infty}^\infty \Sigma(x,y)[{\bar{v}}_{los}(x,y) - v_{sys}]dx }$ = $<$V$>$  and\\
\noindent ${\int\limits_{-\infty}^\infty \Sigma(x,y)x dx}$ = $<$X$>$ \\
Equation (1a) then reduces to
$$ \Omega_p \sin(i) = \frac {<V>}{<X>}                                              \eqno (1b)$$ \\ 
where $<$V$>$ and $<$X$>$ are known as the Tremaine-Weinberg (TW) integrals,
each of which can be computed for different values of $y$.
If the bar has a fixed pattern speed, one could expect to see a fixed 
slope (i.e., $\Omega_p$ sin(i)) in a plot of $<$V$>$ versus $<$X$>$.

\section{Application of Tremaine-Weinberg Method to NGC 3741}

\subsection{NGC 3741}

NGC 3741 is a gas-rich, nearby dwarf irregular galaxy.
\mbox{H\,{\sc i}} 21cm radio synthesis images of this galaxy were obtained from the Giant Meterwave Radio Telescope (GMRT) observations (Begum et al. 2005), as part of the FIGGS survey (Begum et al. 2008b). 
The  \mbox{H\,{\sc i}} mass derived from these observations is $\sim$ 10$^8$ M$_{\odot}$ (Begum et al. 2008a) which is $\sim$ 10 times more than the stellar mass  ($\sim$ 10$^7$ M$_{\odot}$, Begum et al. 2008a). This \mbox{H\,{\sc i}} is in a giant disk, which extends upto 8.3 times the Holmberg radius (Begum et al. 2005). The \mbox{H\,{\sc i}} disk is also unusual in that it shows significant structures i.e. both a bar and spiral arms. As noted in \S 1, it is one of the few galaxies with a purely \mbox{H\,{\sc i}} bar and a spiral arm.
In most spiral galaxies, bars are generally found to be poor in \mbox{H\,{\sc i}} (Huttemeisteret al. 1999).  Interestingly dynamical models of NGC 3741 indicate that  it is one of the most dark matter-dominated dwarf irregulars known with a dynamical mass to light ratio of 107 (Begum et al. 2005). Unlike ordinary large spirals, the dark matter plays a dominant role in NGC 3741 even within the baryonic disk,  and therefore it is an ideal site to test the theoretical models of the dynamical effect of the dark matter halo on the galactic disk,  including the role of dark matter dynamical friction on the bar pattern speed. 

For the analyses here we use the \mbox{H\,{\sc i}} 21cm radio synthesis images of this galaxy obtained from VLA-ANGST (Very Large Array - 
ACS Nearby Galaxy Survey Treasury) survey (Ott et al. 2012).  In Figure 1, we present the \mbox{H\,{\sc i}} emission map (7.6\arcs $\times$ 
6.2\arcs resolution) overlaid with the position of the slits with an assumed PA = 41$^0$ [Left panel] and velocity field of NGC 3741 
[Right panel]. The small, dark, vertically-extended structure seen in Figure 1[Left]
and also the distortions in the central region in the \mbox{H\,{\sc i}} velocity map of NGC 3741 (Figure 1[Right]) are indicative of the 
presence of a bar. The different parameters of NGC 3741 have been summarized in Table 1.

\begin{table}
\caption{Parameters for NGC~3741}
\label{tab:par}
\vskip 0.1in
\begin{tabular}{ll}
\hline
Parameters& Value \\
\hline
\hline
Distance & 3.24 Mpc $^{(1)}$ \\
$v_{sys}$ & 229.1 kms$^{-1}$ $^{(1)}$\\
${\rm{M_{gas}}~(1.3M_{HI})}$& 1.69 $\times~10^8$~M$_\odot$ $^{(2)}$\\
${\rm{L_B}}$& 2.7 $\times~10^7$~L$_\odot$ $^{(2)}$\\
${\rm{M_{gas}/L_B}}$& 6.26 $^{(2)}$\\
Mass-to-light ratio, $\Gamma_B$&0.51 $^{(2)}$ \\
Stellar mass ${\rm{M_*}}$& 1.38 $\times~10^7$~M$_\odot$ $^{(2)}$ \\
${\rm{M_{gas}/M_*}}$& 12.2 $^{(2)}$\\
Total dynamical mass (${\rm{M_T}}$)&4.03 $\times~10^9$~M$_\odot$ $^{(2)}$ \\
${\rm{M_T/L_B}}$& 149 $^{(2)}$\\
${\rm{M_{dark}/M_{T}}}$(\%)& 95 $^{(2)}$ \\
\mbox{H\,{\sc i}} diameter & 13.8 $'$ $^{(3)}$ \\
B band Holmberg diameter & 100 $''$ $^{(3)}$ \\
DM (pseudo-isothermal) core density ${\rho}_{0}$ & 0.078 M$_\odot$pc$^{-3}$ $^{(3)}$ \\
DM (pseudo-isothermal) core radius $r_c$ & 0.7 kpc $^{(3)}$ \\
Rotational velocity $v_{rot}$ & 44 kms$^{-1}$ $^{(3)}$ \\
Position Angle of Bar & $\sim -8.0^{o}$           \\
\hline
\footnote*{} Ott et al. (2012) \\
\footnote*{} Begum et al. (2008a) \\
\footnote*{} Begum et al. (2005) \\

\end{tabular}
\end{table}

\begin{figure*}
\begin{center}
\begin{tabular}{cc}
\resizebox{85mm}{!}{\includegraphics{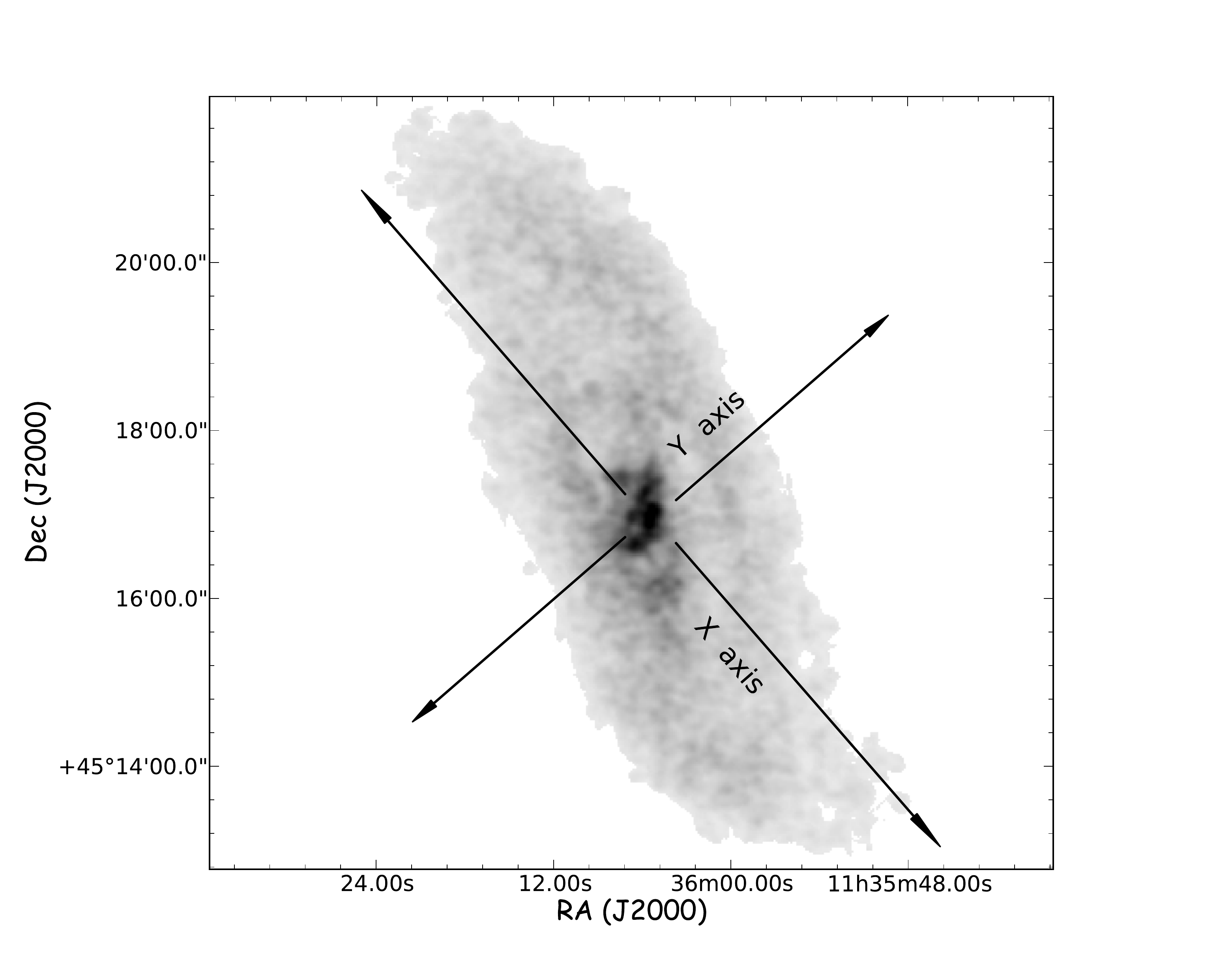}} &
\resizebox{88mm}{!}{\includegraphics{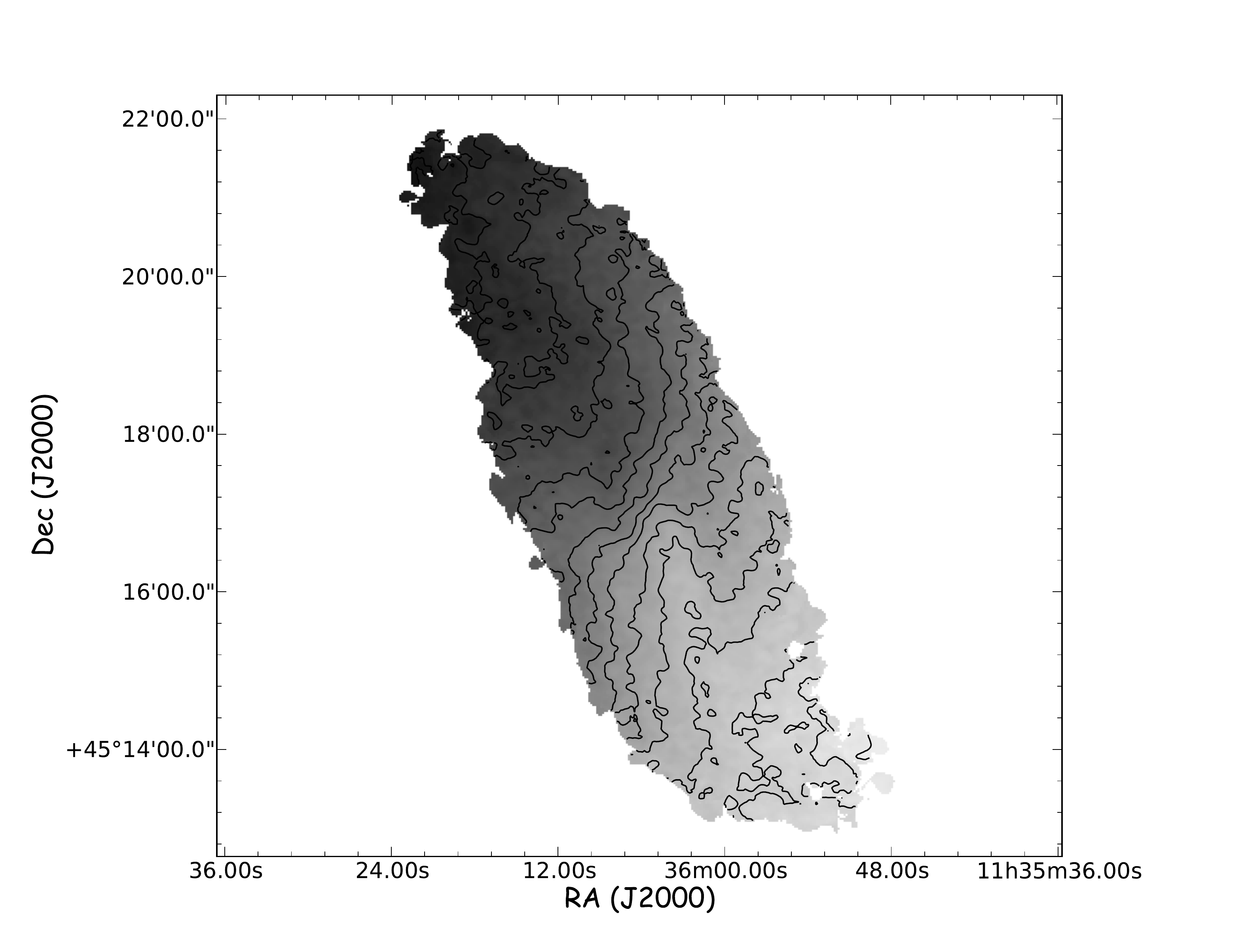}} \\
\end{tabular}
\end{center}
\caption{The V(ery) L(arge) A(rray) Images of NGC 3741 of the VLA-ANGST (Very Large Array - ACS Nearby Galaxy Survey Treasury) survey: 
[Left]: The VLA 7.6"$\times$ 6.2" resolution integrated \mbox{H\,{\sc i}}
emission map of NGC 3741. Overlaid on the plot are the major and minor axes of the galaxy indicated as X and Y axes respectively, assuming the position angle (PA) of the galaxy to be 41$^o$. The small, 
dark, vertically-extended structure seen here is indicative of a bar. Also the two spiral arms of the galaxy are seen as well. [Right]: 
The \mbox{H\,{\sc i}} velocity field of NGC 3741, the contours being in the steps 
of 6.7 kms$^{-1}$ and range from $\sim$ 174 kms$^{-1}$ to $\sim$ 281 kms$^{-1}$. The distortions in the central region in the \mbox{H\,{\sc i}} velocity map indicates the presence of a bar. }
\label{fig1}
\end{figure*}

\subsection{\mbox{H\,{\sc i}} as a tracer of pattern speed}

In the TW method, the main requirement for a population to be the tracer of pattern speed is that its 
surface brightness remains unchanged with
 time. The dust-free old stellar population is therefore 
an ideal choice for a tracer. In general, \mbox{H\,{\sc i}} is believed to not be a good tracer as its intensity 
may vary with time due to its possible ionization, conversion to molecular gas and the variations 
in optical depth. In this particular case, however, these concerns are less serious than for \mbox{H\,{\sc i}} in typical large 
spirals. 
This is because ionization of \mbox{H\,{\sc i}} is negligible in  
NGC 3741 as the star formation rate is very low, 3.4$\times$10$^{-3}$ M$_{\odot}$yr$^{-1}$ (Begum et al. 2008a, Roychowdhury et al. 2009)
and therefore the source of ionizing radiation minimal. 
Besides the molecular gas content ($H_2$) of dwarf galaxies is small (Israel 1995, Taylor et al. 1998) , indicating that conversion of the \mbox{H\,{\sc i}} to the molecular phase is also not important.

\section{Calculations \&  Results}
\label{result}

In Figure 2, we plot each of the TW integrals $<$X$>$ and $<$V$>$ as a function of $y$ which is the distance of the slit from the line-of-nodes [Left-panels]. We also plot $<$V$>$ as a function of $<$X$>$ and a linear fit to the data [Right panel]. The centre of the galaxy was chosen such that $<$V$>$ versus $<$X$>$ plot is most symmetric about $<$X$>$ = 0. It may be noted this kinematic centre [RA: 11h 36m 7.07s, DEC: 45d 17m 13.96s] is 20" away from 
the photometric center as obtained from NED [RA: 11h 36m 5.743s, DEC: 45d 16m 59.96s]. The TW method assumes that all the input parameters in Equation (1d) remain constant over the entire disk of the galaxy. In NGC 3741, however, as indicated by the tilted-ring analysis, the position angle PA and the inclination ``$i$" vary with radius 
(Begum et al. 2005). A wrong value of the PA can be a potential source of error in the TW method (Debattista 2003). 
We calculate the TW integrals $<$V$>$ and $<$X$>$ for a range of values of the PA and find that the pattern speed 
${\Omega}_p$ systematically increases with assumed PA at the rate of $\sim$ 4$\%$  per degree, consistent with similar 
findings in earlier studies (Debattista 2003, Debattista \& Williams 2004, Chemin \& Hernandez 2009). The value adopted 
here corresponds to the PA for which we get the best linear fit (as determined by the minimum of the reduced chisquare) 
to the $<$V$>$ versus $<$X$>$ plot, namely PA = 41$^o$. We use the uncertainty in the position angle, which is 
$\sim 5^{o}$ in the bar region, to derive the uncertainty in the pattern speed. The error in the pattern speed is dominated 
by this uncertainty in the structural parameters of the galaxy.  We also note that the $<$V$>$ versus $<$X$>$ curve remains 
roughly linear for -50\arcs $<$ $<$X$>$ $<$ 50\arcs, indicating a constant pattern speed exists only within the corresponding 
range of offset ($y$) in this  galaxy. Interestingly, the above $<$X$>$ range corresponds to  -47\arcs $<$ $y$ $<$ 27\arcs, which 
is the region containing the bar only. In other words, the constant value of the pattern speed obtained here corresponds to only 
the bar and not the spiral arms. Assuming an average inclination $i  = 55^o$ (Begum et al. 2005), the bar pattern speed 
$\Omega_p = 0.27 \pm 0.05 \textrm{kms}^{-1}{\textrm{arcsec}}^{-1}$ (17.1$\pm $ 3.4\textrm{kms}$^{-1}{\textrm{kpc}}^{-1}$ 
for distance to the galaxy D = $3.24$ Mpc, see for example Ott et al. 2012).
\begin{figure}
\centering
\includegraphics[angle=0, width=0.5\textwidth]{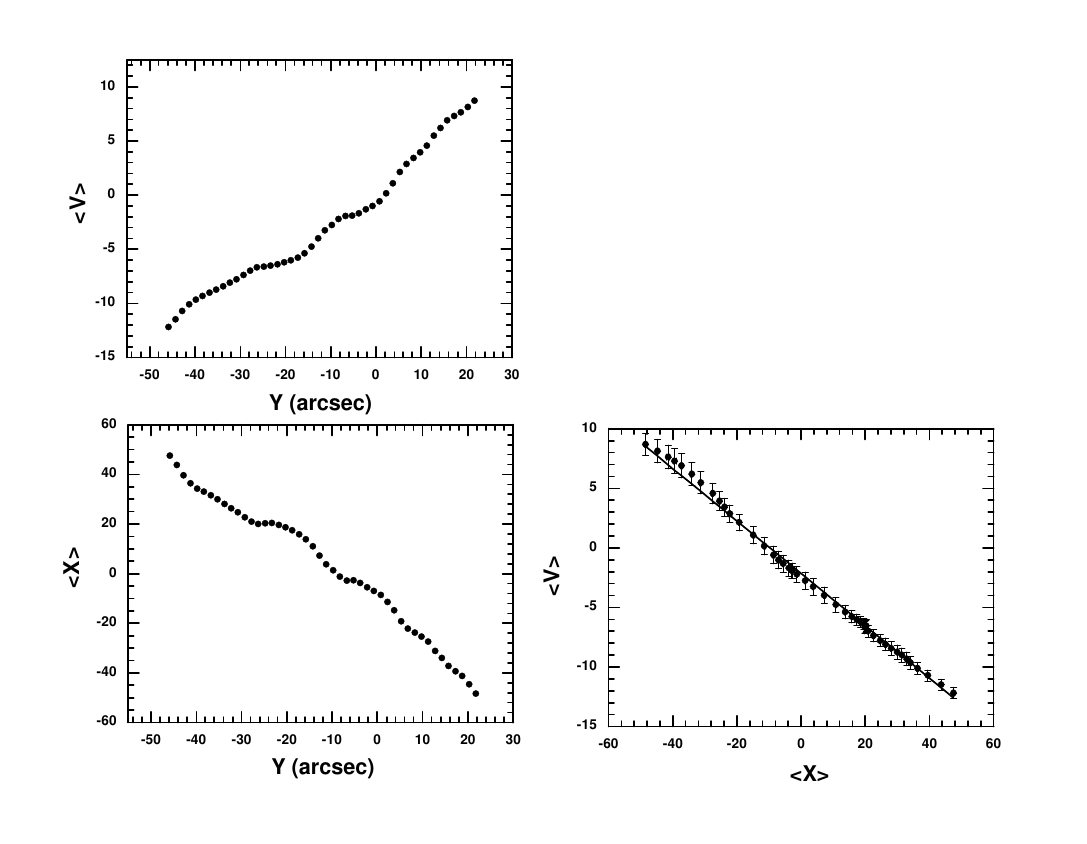}
\caption{Pattern speed measurement of the dwarf irregular galaxy NGC 3741: 
Here we present the Tremaine-Weinberg integrals for position angle PA =  41$^o$. On the left are shown $<V>$ 
(the integral of surface brightness-weighed radial velocity along the slit) as a function of slit offset y 
(distance of the slit from the line of nodes) and $<X>$ (the integral of surface brightness-weighed position vector along a slit) as a 
function of y. On the right, $<V>$ is plotted against $<X>$ overlaid with the best straight line fit to it within -50\arcs $<$ $<X>$ $<$ 
 50\arcs. The slope (viz. 0.22 km$s^{-1}$arcsec$^{-1}$) of the $<V>$ versus $<X>$ plot gives $\Omega_p$ sin(i) where 
$\Omega_p$ is the pattern speed and ``i" the inclination of the disk.}
\label{fig2}
\end{figure}

\subsection{Estimation of the bar semi-major axis $r_b$ }

A review of the different methods for bar length determination is given in Athanassoula \& Misiriotis (2002).
We use the following three techniques to estimate the bar semi-major axis length $r_b$ in NGC 3741.\\

\noindent \emph{Visual Inspection}: This is the most direct but also the crudest method of bar length estimation which involves visual inspection of the \mbox{H\,{\sc i}} intensity map.
The \mbox{H\,{\sc i}} intensity ($\Sigma_{\mbox{H\,{\sc i}}}$) is plotted as a function of the distance along the major axis from the centre of the map as shown in Figure 3 [Left panel]. 
We take the sky plane projected value of the bar-length $2 r_b$ to be the zone beyond which the intensity drops sharply on either side of the 
centre, which is $\sim$ 101\arcs from
the figure. Assuming the position angle of the bar $\sim$ -8$^o$ and an inclination of the galaxy $i = 55^{o}$, deprojected value 
of the bar length ($2 r_b$) on the galaxy plane is $\sim$ 128\arcs. Assuming $D = 3.24$~kpc, $r_b \sim 1 $~kpc.\\

\noindent \emph{Position Angle Variation}: In this technique, the position angles PAs of the elliptical isophotes fitted to the 
\mbox{H\,{\sc i}} intensity map (obtained using isophote in IRAF) are plotted against 
the major axis length (in arcsec) as in Figure 3 [Middle panel]. A sharp change in position angle is expected be seen just ouside the bar 
region from which the bar semi-major axis ($r_b$) can be estimated. 
The figure indicates the projected value of 2$r_b$ along the major axis in the sky plane is $\sim$ 117\arcs, corresponding to a 
deprojected bar-length (2$r_b$) on the galaxy plane of 148$''$, or $r_b$ $\sim$ 1.2~kpc.\\

\noindent \emph{Fourier Decomposition Method}: 
 The Fourier Method constitutes determination of bar length by decomposing the azimuthal luminosity profiles into the different Fourier modes or components (See, for example, Aguerri et al. 2000). If
I(r, $\theta$) denotes the deprojected luminosity profile, then its Fourier series is given by
$$ I(r, \theta) =  A_0(r)/2 + \Sigma ( A_m(r)cos(m \theta) + B_m(r)sin(m \theta) ) \eqno 2(a) $$
where the coefficients are given by
$$ A_m(r) = 1/\pi \int_0^{2\pi} I(r, \theta)cos(m\theta)d\theta \eqno 2(b) $$
and
$$ B_m(r) = 1/\pi \int_0^{2\pi} I(r, \theta)sin(m\theta)d\theta  \eqno 2(c) $$
The amplitude of the m$^{th}$ Fourier component is thus given by
$$ I_0(r) = A_0(r)/2  \eqno 2(d) $$
and
$$ I_m(r) =  \sqrt {A^{2}_{m}(r) + B^{2}_{m}(r)}  \eqno 2(e) $$
In the region where the bar is located, the even modes predominate with the m = 2 the leading component. 
According to Ohta et al. (1990), the intensity of the bar is given by
$$ I_b = I_0 + I_2 + I_4 + I_6 \eqno 2(f) $$
and the interbar intensity is
$$ I_{ib} = I_0 - I_2 + I_4 - I_6  \eqno 2(g) $$
The radius of the bar $r_b$ is then determined by the outer radius where the straight line 
$$ (I_b/I_{ib}) = \frac{ (I_b/I_{ib})_{max} - (I_b/I_{ib})_{min} }{2} + (I_b/I_{ib})_{min} \eqno 2(h)$$ 
intersects the $I_b/I_{ib}$ vs R curve
In Figure 3 [Right Panel], we use this method to determine the bar length 2$r_b$ on the galaxy plane to be $\sim$ 149 $''$, which gives 
$r_b$ $\sim$ 1.2~ kpc.\\  

\noindent We adopt $r_b$ = 1.2 kpc for the rest of our calculations as this is also a conservative value to adopt, 
in view of the conclusion below that the bar is slow.

\begin{figure*}
\begin{center}
\begin{tabular}{ccc}
\resizebox{50mm}{!}{\includegraphics{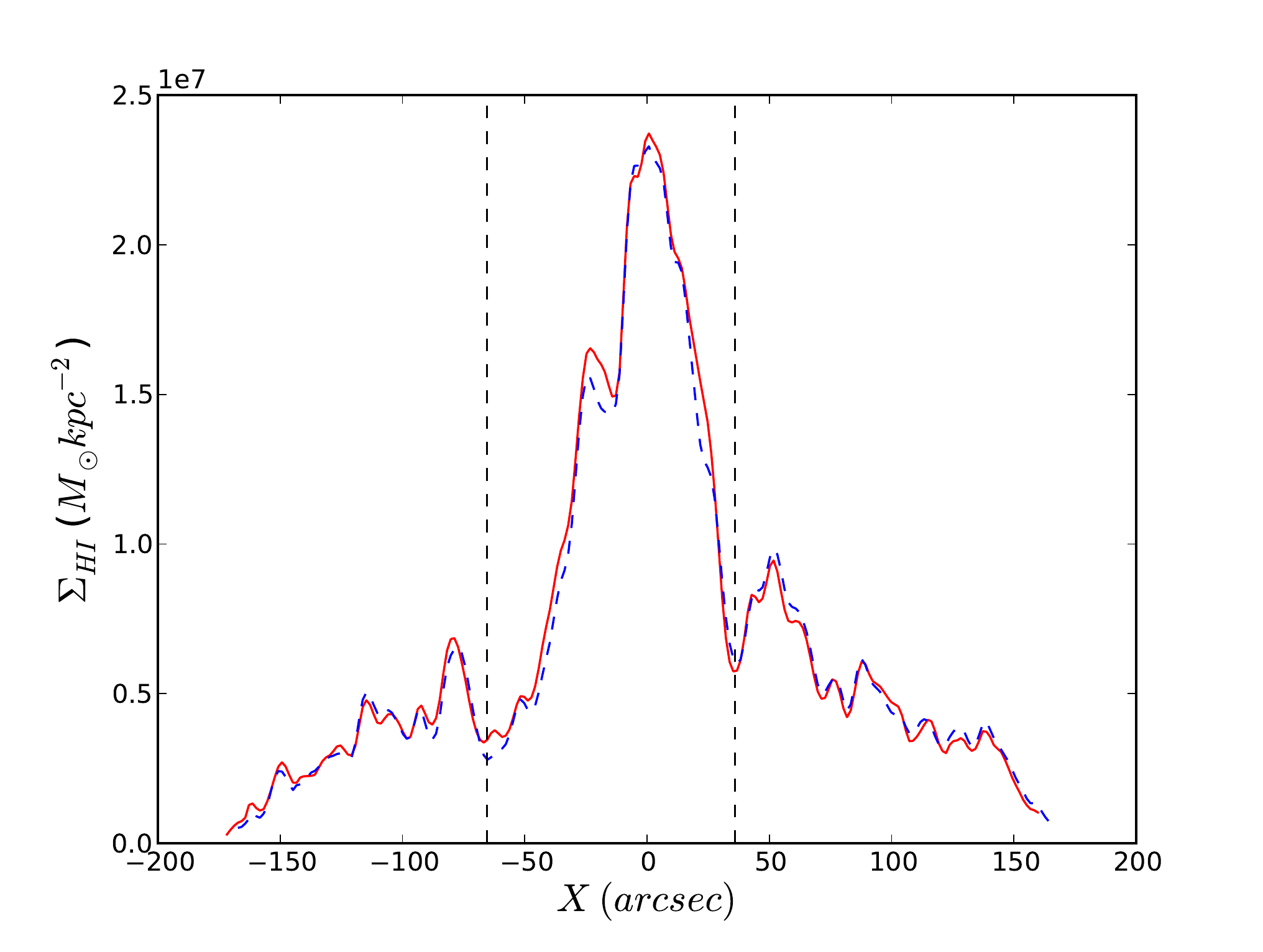}} &
\resizebox{50mm}{!}{\includegraphics{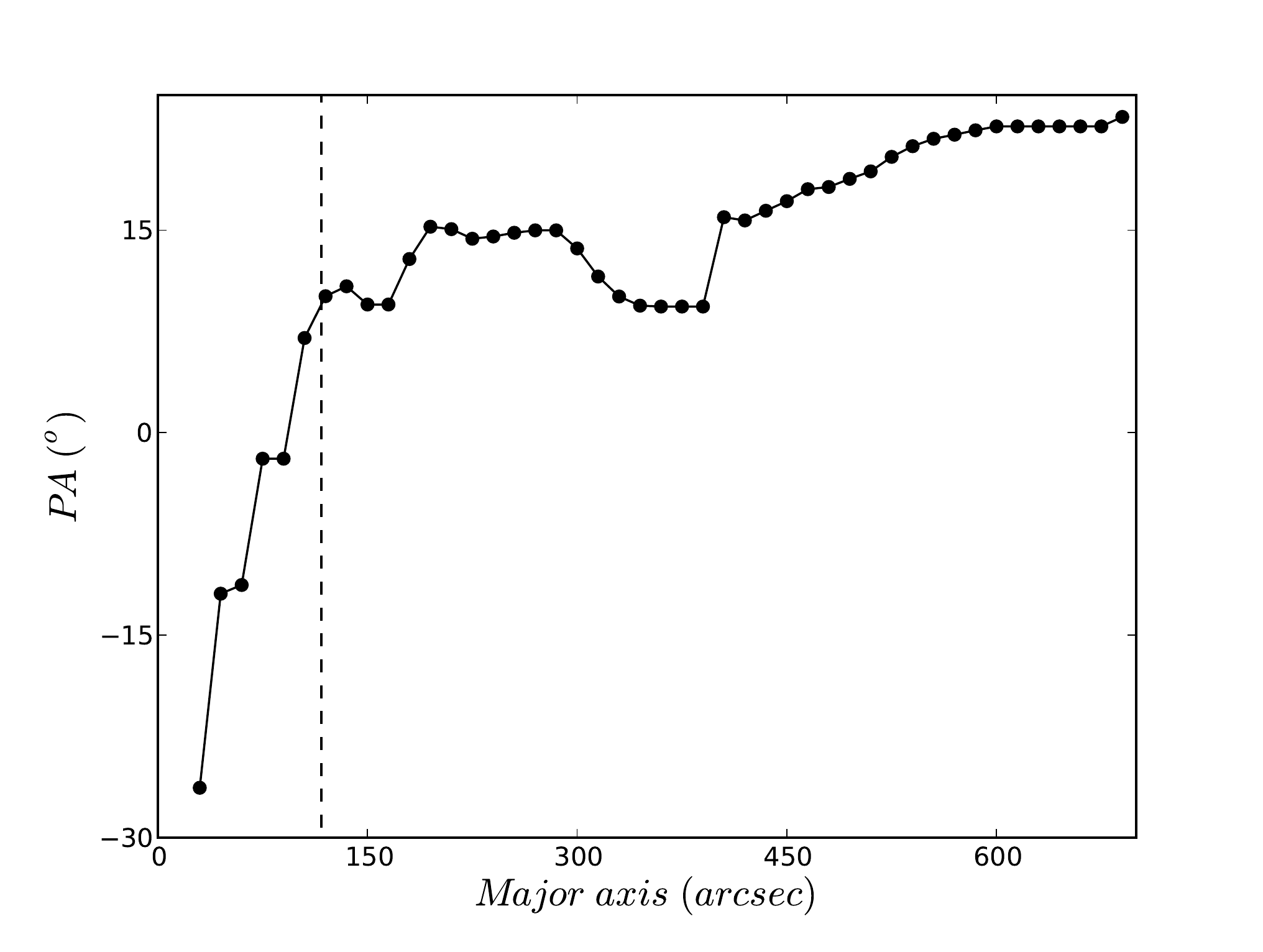}} &
\resizebox{50mm}{!}{\includegraphics{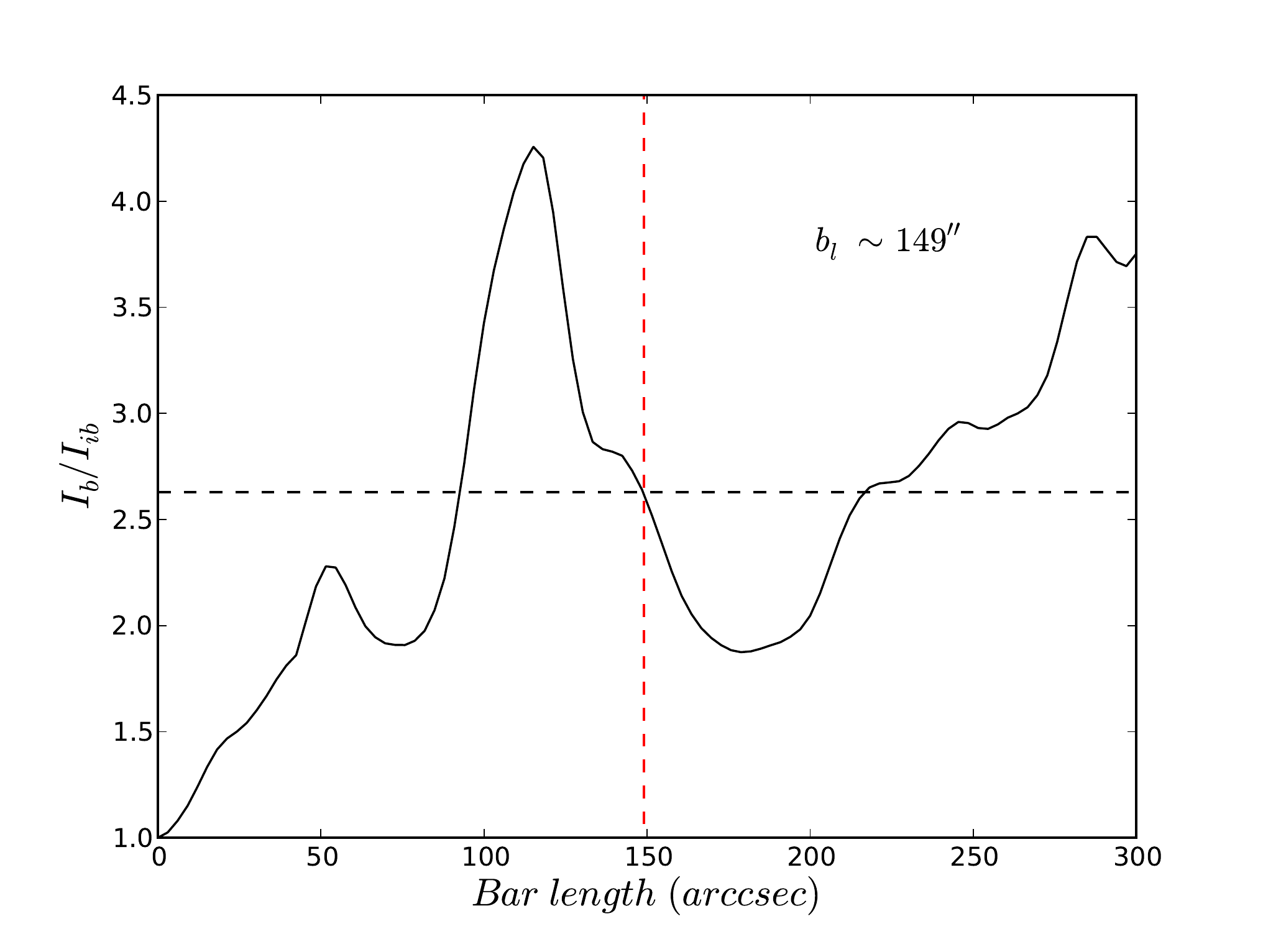}} \\
\end{tabular}
\end{center}
\caption{Estimation of bar length of NGC 3741 by three different methods:
[Left] \emph{Visual Inspection Method:} Here the intensity is plotted as a function of the distance along the major axis. 
The bar-length (2 $r_b$) is given by the zone beyond which intensity drops sharply on either side of the centre, which is $\sim$ 101\arcs  
from the figure. Assuming the position angle of the bar $\sim$ -8$^o$, an inclination of the galaxy $i = 55^{o}$ and distance to the galaxy 
$D = 3.24 kpc$,  deprojected value of the bar radius $r_b$ on the galaxy plane is $\sim 1$~kpc. 
[Middle] \emph{Position Angle Method:} We plot the position angles (PAs) of isophotes as a function of distance along the major axis from 
the centre (in arcsec), and the value at which there is an abrupt change in the value of the
position angle gives the projected value of the bar length (2 $r_b$) on the sky plane, which is $\sim$ 117\arcs as seen in this figure. 
The deprojected value of $r_b$ by this method is $r_b \sim 1.2$~kpc. 
[Right] \emph{Fourier Decomposition Method:} In this method, we plot $I_b/I_{ib}$, the ratio of the bar intensity to the interbar intensity, 
as a function of the distance along the length of the bar (in arcsec).
The bar-length $ 2 r_b$ is then determined by the outer radius where the straight line
$ (I_b/I_{ib}) = \frac{ (I_b/I_{ib})_{max} - (I_b/I_{ib})_{min} }{2} + (I_b/I_{ib})_{min} $
intersects the $I_b/I_{ib}$ vs R curve, which, according to the figure, is $\sim$ 149\arcs, or $r_b \sim$~1.2 kpc.}
\label{fig3}
\end{figure*}

\subsection{The ratio $\cal R$ of corotation radius $R_{cr}$ to bar semi-major axis $r_b$}

The ratio $\cal R$ of the corotation radius $R_{cr}$ (the radius at which the pattern speed equals the rotational velocity of the galaxy) to the bar semi-major axis $r_b$ is the parameter conventionally used 
to determine if the pattern speed in a galaxy is ``slow'' or ``fast'', and also to compare the pattern speeds of different galaxies. 
Theoretical calculations of orbits in a bar plus disk potential have shown that the self-consistency of bars requires $\cal R \geq $ 1. Bars are defined as ``fast" if $1 \leq \cal R \leq $ 1.4 
and ``slow" if $ \cal R \geq $ 1.4 (See, for example, Debattista 2003). For NGC 3741, 
$R_{cr}$, as determined from the observed rotation curve, is $R_{cr}$ = 1.8 $\pm$ 0.4 kpc.
With our adopted value of $r_b$ $\sim$ $1.2$ kpc, the ratio $\cal R$ = ${R_{cr}}/{r_b}$ = 1.6 $\pm$ 0.3, 
indicating that the bar in NGC  3741 is slow, since NGC 3741 is dominated by dark matter at all radii, including the inner disk where the 
bar is located (\S 3.1) unlike in large spiral galaxies (Banerjee \& Jog 2008 and references therein), therefore our result is consistent 
with the fact that the dynamical friction due to the dark matter halo slows down the bar, as predicted earlier by analytical (Weinberg 1985) 
and galaxy simulation studies (Debattista \& Sellwood 1998, 2000).

\section{Discussion}

\subsection*{The slow bar in NGC 3741 and its dense dark matter halo} The dwarf irregular galaxy NGC 3741 is dominated by dark matter at all 
radii as is evident from its rotation curve studies. Begum et al. (2005) modeled the observed rotation curve of the galaxy using both a pseudo-isothermal (Binney \& Tremaine 2008) and an NFW (Navarro et al. 1996) dark matter halo. A best-fit pseudo-isothermal halo gave a core 
radius $r_c$ = 0.7 $\pm$ 0.1 kpc and core density $\rho_0$ = 77.9 $\pm$ 8.9 $\times$ $10^{-3}$ \textrm{$M_\odot$} pc$^{-3}$. The best fit NFW mass model gave a concentration parameter c = 11.4 $\pm$ 0.8 and $V_{200}$ = 35.3 $\pm$ 1.4 km$s^{-1}$. The fits were equally good for both the models. Also, in both the cases,  modeling of the rotation curve indicated, that dark matter dominates the galactic disk from the innermost radii. Dynamical friction due to a dense dark matter halo is expected to slow down a rotating bar (see for example, Chandrasekhar 1943), which was also indicated both by analytical studies and galaxy N-body simulations (\S 4.2). Therefore our detection of a slow bar in the dark matter-dominated dwarf irregular galaxy NGC 3741 is consistent with the earlier theoretical predictions. Interestingly, the two other dark matter-dominated galaxies to which TW method has been applied, NGC 2915 and UGC 628, are also found to host slow bars. For NGC 2915, $\cal R$ = 1.7 (Bureau et al. 1999) while for UGC 628, $\cal R$ = 2.$^{+0.5}_{-0.3}$ (Chemin \& Hernandez 2009). 

\subsection*{The pattern speed of bar and the spiral arm in NGC 3741} 

	The \mbox{H\,{\sc i}} intensity map and velocity field of NGC 3741 also bear signatures of the presence of a spiral pattern in addition to the central bar. However the TW method does not yield a constant pattern speed for both the bar and the spiral arm as was discussed in \S 3. This is in contrast to the observations in NGC 2915 in which the \mbox{H\,{\sc i}}-rich bar and a spiral arm were found to share a constant pattern speed (Bureau et al. 1999).	We note that Sellwood \& Sparke (1988) suggests that spiral arms in barred spiral galaxies could have a different (lower) pattern speed than the bar.
It is also possible that the spiral arm has a pattern speed systematically varying with radius (Westpfahl 1998; Merrifield et al. 2006; Meidt et al. 2008; Speights \& Westpfahl 2011). In fact, Speights \& Westpahl (2012) find evidence for radially-varying pattern speed in the dwarf galaxies NGC 3031, NGC 2366 \& DDO 154. In NGC 3031, the pattern speed is equal to the material speed at all radii. In NGC 2366 and DDO 154, however, the pattern speed shows deviations from the material speed at places, with the angular resolution of the data showing a prominent effect in the radial-variation of the pattern speed studies. 

\subsection*{Correlation between $\cal R$ and dark matter dynamical friction} We also study the correlation, if any, between the pattern speed and the dark matter dynamical friction in a galaxy. While the pattern speed is well-parametrized by $\cal R$ (\S 4.2), getting a closed form expression for the dynamical friction felt by the bar due to its motion through the dark matter halo is non-trivial. We therefore approximate it by the analytical expression for the dynamical friction on a rigid body in motion through a cloud of particles (Chandrasekhar 1943). According to this expression, dynamical friction varies directly as the density of the particles and inversely as the velocity cube of the rigid body. In Figure 4[Left], we plot $\cal R$ versus log $({\rho}_0/{v_{rot}}^3)$ (where ${\rho}_0$ is the dark matter core density in a pseudo-isothermal halo and 
$v_{rot}$ is the rotational velocity of the galaxy) for different galaxies for which pattern speed have been determined (See Table 2). 
For this, we had to restrict ourselves to the galaxies for which both the bar pattern speed has been measured and the dark matter density 
profile modeled from observations. The figure provides shows the dependence of $\cal R$ on $\rho_0/{v_{rot}}^3$  with their correlation 
coefficient being -0.37. In Figure 4[Right], we separately check for the correlation between $\cal R$ and ${\rho}_0$, the value
of the correlation coefficient is -0.32. Note that the pattern speed of NGC 925 has been obtained from kinematical arguments and not 
using TW method like for the other galaxies in this plot. Excluding NGC 925 from the sample makes the correlation between 
$\cal R$ on $\rho_0/{v_{rot}}^3$  tighter (0.98), but the number of data points is only 4. This study hence provides modest support 
for the idea that dynamical friction plays a role in determining the speed of bars but a larger sample is needed to reach a firm conclusion.

\begin{table*}


   \caption{List of galaxies for which the correlation between $\cal R$ and ${\rho}_0/{v_{rot}}^3$ is studied}
   \label{table2}
\begin{tabular}{llllll}
\hline
Galaxy   & $\rho_{0}$ & $V_{\rm rot}$ & $R_{\rm CR}$ & ${r_{\rm b}}$ & References \\
         & ($M_{\odot}pc^{-3}$) & (kms$^{-1}$) & (\arcs) & (\arcs) &  \\
\hline

NGC 925  & 0.003 $\pm$ 0.001 & 120 $\pm$ 10 & 374 & 120 & Elmegreen et al. (1998),\\
         & & & & & de Blok et al. (2008) \\
NGC 4736 & 0.022 $\pm$ 0.041& 120 $\pm$ 10 & 26 &20 & Rand \& Wallin (2004), \\
         & & & & & de Blok et al.(2008) \\
NGC 628  & 0.152 $\pm$ 0.028 & 110 $\pm$ 5& $28.6^{+8.5}_{-5.8}$ & 14.0 $\pm$ 0.8 & Chemin \& Hernandez (2009), \\
         & & & & & de Blok \& Bosma (2002) \\
NGC 2915 & 0.10 $\pm$ 0.02 & 82 & $390^{+\infty}_{-80}$ & 180& Bureau et al. (1999), \\
NGC 3741 & 0.078 $\pm$ 0.009& 44 $\pm$ 1.3& 116.8 $\pm$ 1.69 & 74.5 & Current paper \\
         & & &  & & Begum et al. (2005) \\
\hline

\footnote{} 

\end{tabular}


\end{table*}

\begin{figure*}
\begin{center}
\begin{tabular}{ccc}
\resizebox{85 mm}{!}{\includegraphics{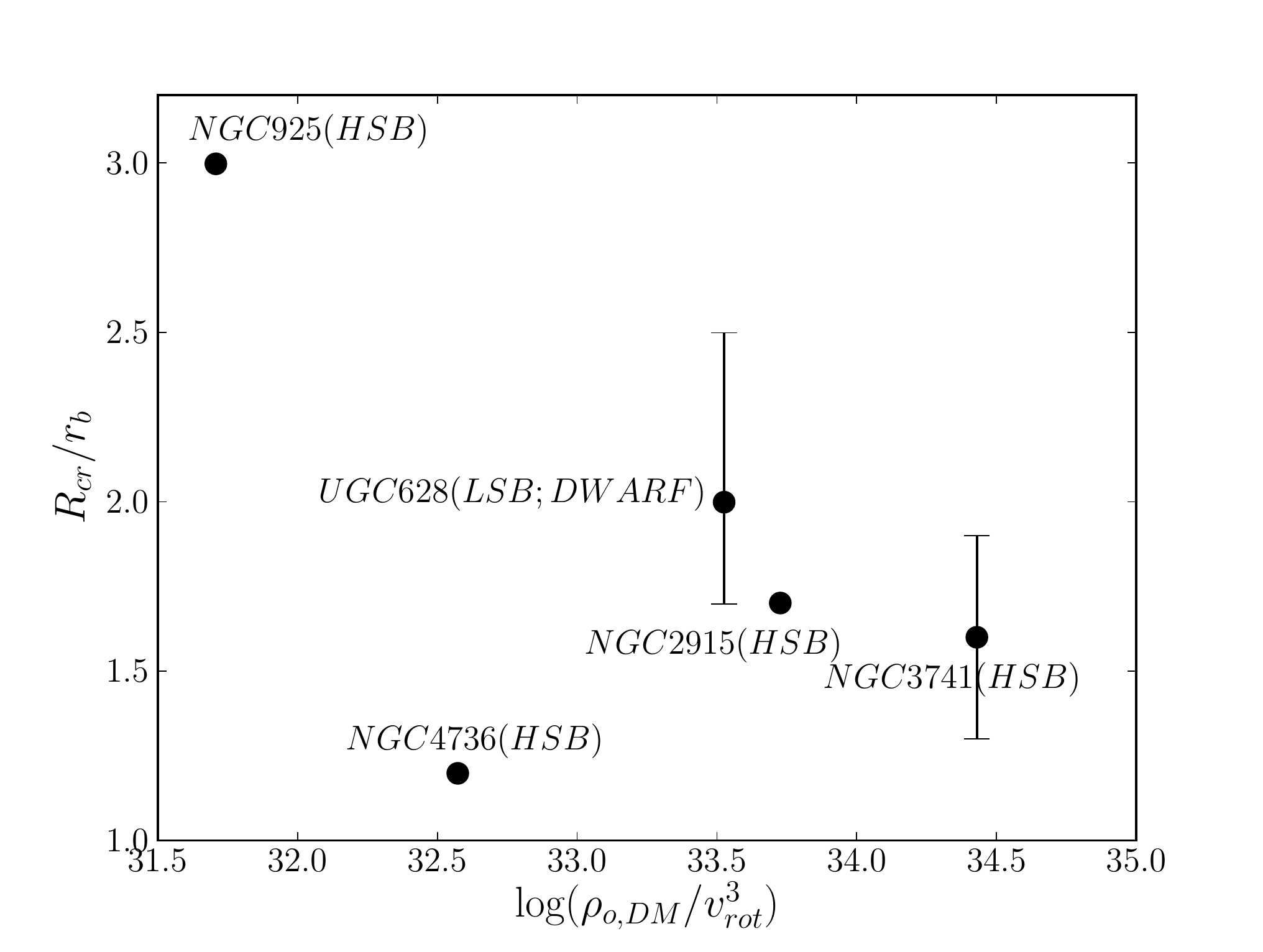}} &
\resizebox{85 mm}{!}{\includegraphics{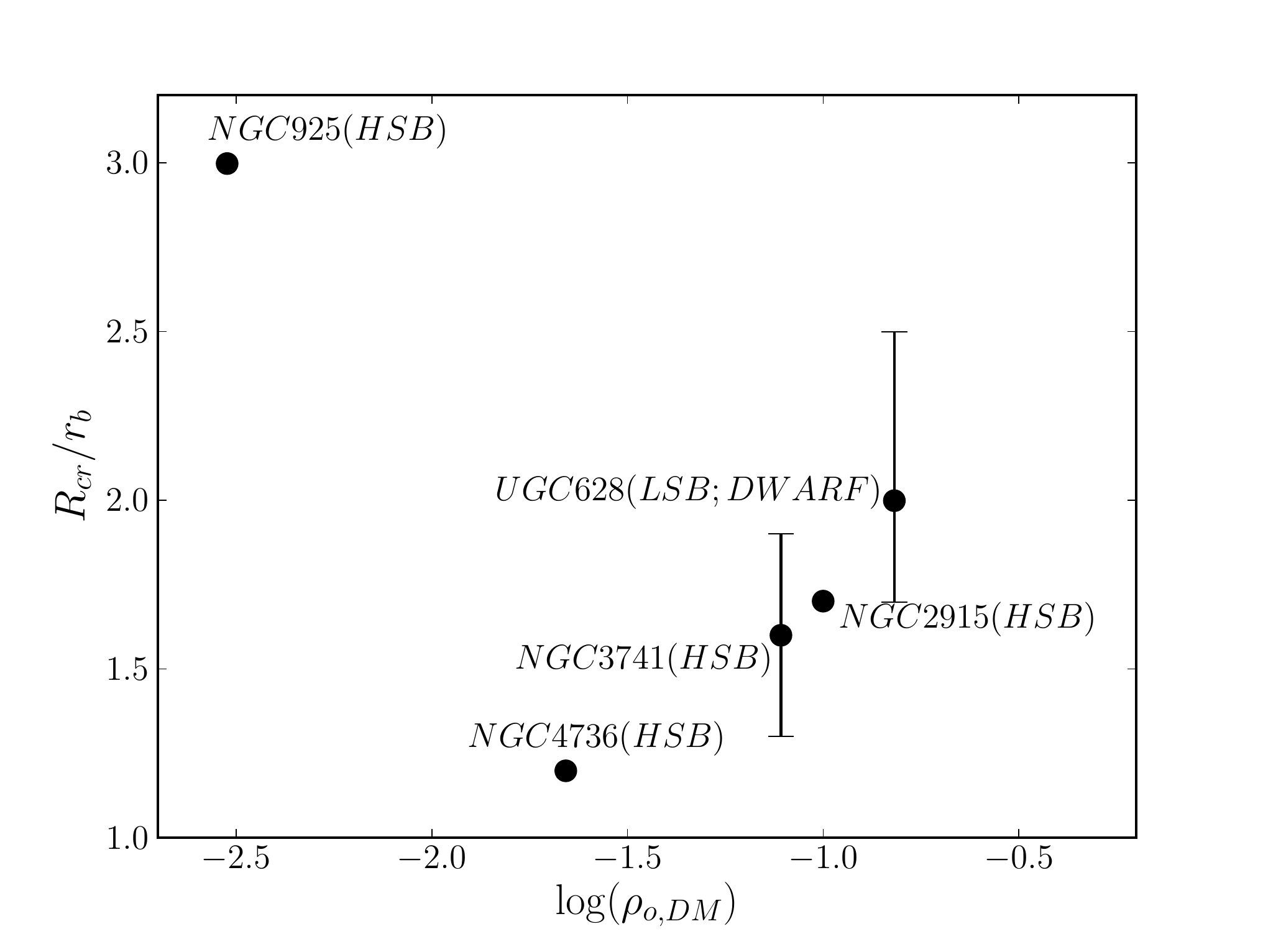}} \\
\end{tabular}
\end{center}
\caption
{[Left] Plot of $R_{cr}/{r_b}$ (= $\cal R$), the ratio of the corotation radius to the bar semi-major axis length, against 
log($\rho_0/{v_{rot}}^3$),
where $\rho_0$ is the core density of the pseudo-isothermal dark matter halo and $v_{rot}$ is the rotational velocity of the galaxy.
Large values of $\cal R$ indicate a slow bar 
and vice versa. The dynamical friction due to the dark matter halo roughly scales as $\rho_0/v_{rot}^3$. 
The correlation coefficient between $\cal R$ and $\rho_0/{v_{rot}}^3$ from this plot is -0.37. 
Note that error bars on $\cal R$ were available only for UGC 628 and NGC 3741.
[Right] Plot of $\cal R$ versus log($\rho_0$) to study the dependence of the bar pattern speed on the dark matter density only. Here the correlation coefficient 
is -0.32.} 
\label{fig4}
\end{figure*}

\section{Conclusions}

Applying the Tremaine-Weinberg method, we obtain the pattern speed in NGC 3741, using neutral hydrogen \mbox{H\,{\sc i}} as tracer as 
observed in the 21 cm radio-synthesis data from the VLA-ANGST survey. NGC 3741 is a dark matter dominated dwarf irregular galaxy with a 
large \mbox{H\,{\sc i}} disk,  as well as a bar and a spiral arm rich in \mbox{H\,{\sc i}}, which is a rare feature as bars appear to be gas 
poor in general. We measure a constant pattern speed of $\Omega_p$ = 17.1 $\pm$ 3.4 kms$^{-1}$kpc$^{-1}$ which corresponds to 
only the bar and not the spiral arms.
Also the ratio of the corotation radius to the bar semi-major axis is $\sim 1.6$ indicating a slow bar. This is possibly the effect of the 
dynamical friction due to the dense dark matter halo as predicted  
earlier by theoretical and N-body simulation studies. Our detection of a slow bar in NGC 3741 is the first detection of its kind 
in a dark matter dominated dwarf irregular galaxy.

\end{document}